\documentclass{emulateapj}
\usepackage{apjfonts}
\usepackage{rotating}

\newcommand{\pivec}{\mbox{\boldmath $\pi$}}

\lefthead{SHIN ET AL.} 
\righthead{CHARACTERIZING MICROLENSING BINARIES}

\begin{document}
\title{Characterizing Low-Mass Binaries From Observation of Long Time-scale Caustic-crossing  
Gravitational Microlensing Events}

\author{
I.-G. Shin$^{1,G3}$,
C. Han$^{1,S,G3}$,
J.-Y. Choi$^{1,G3}$,
A. Udalski$^{O01,G1}$, 
T. Sumi$^{M01,G2}$,        
A. Gould$^{u01,G3}$,
V. Bozza $^{702,722,G5}$,
M. Dominik $^{703,\star,G5}$,
P. Fouqu\'e$^{P01,G6}$,
K. Horne$^{703,G4}$,\\ 
and\\
% OGLE -----------------------------
M.\,K. Szyma{\'n}ski$^{O01}$,
M. Kubiak$^{O01}$, 
I. Soszy{\'n}ski$^{O01}$,
G. Pietrzy{\'n}ski$^{O01,O02}$, 
R. Poleski$^{O01}$, 
K. Ulaczyk$^{O01}$, 
P. Pietrukowicz$^{O01}$,
S. Koz{\l}owski$^{O01}$,
J. Skowron$^{u01}$,
{\L}. Wyrzykowski$^{O01,O03}$\\ 
(The OGLE Collaboration),\\
% MOA ------------------------------
F. Abe$^{M02}$,
D.P. Bennett$^{M03}$,
I.A. Bond$^{M04}$,
C.S. Botzler$^{M05}$,
P. Chote$^{M06}$,
M. Freeman$^{M05}$,
A. Fukui$^{M07}$,
K. Furusawa$^{M02}$,
Y. Itow$^{M02}$,
S. Kobara$^{M02}$,
C.H. Ling$^{M04}$
K. Masuda$^{M02}$,
Y. Matsubara$^{M02}$,
N. Miyake$^{M02}$,
Y. Muraki$^{M02}$,
K. Ohmori$^{M02}$,
K. Ohnishi$^{M08}$,
N.J. Rattenbury$^{M05}$,
To. Saito$^{M10}$,
D.J. Sullivan$^{M06}$,
D. Suzuki$^{M01}$,
K. Suzuki$^{M02}$,
W.L. Sweatman$^{M04}$,
S. Takino$^{M02}$,
P.J. Tristram$^{M06}$,
K. Wada$^{M01}$,
P.C.M. Yock$^{M05}$\\
(The MOA Collaboration),\\
% RoboNet --------------------------
D.M. Bramich$^{R04}$,
C. Snodgrass$^{R06}$,
I.A. Steele$^{R05}$, 
R.A. Street$^{R02}$, 
Y. Tsapras$^{R02,R03}$\\ 
(The RoboNet Collaboration),\\
% MiNDSTEp -------------------------
K.A. Alsubai$^{701}$,
P. Browne$^{703}$,
M.J. Burgdorf$^{704,705}$,
S. Calchi Novati$^{702, 706}$,
P. Dodds$^{703}$,
S. Dreizler$^{707}$,
X.-S. Fang$^{708}$,
F. Grundahl$^{709}$,
C.-H. Gu$^{708}$,
S. Hardis$^{710}$,
K. Harps{\o}e $^{710,711}$,
T.C. Hinse$^{710,712,713}$,
A. Hornstrup$^{714}$,
M. Hundertmark$^{703,707}$,
J. Jessen-Hansen$^{709}$,
U.G. J{\o}rgensen$^{710,711}$,
N. Kains$^{715}$,
E. Kerins$^{716}$,
C. Liebig$^{703}$,
M. Lund$^{709}$,
M. Lunkkvist$^{709}$,
L. Mancini$^{717}$,
M. Mathiasen$^{710}$,
M.T. Penny$^{716, u01}$,
S. Rahvar$^{719,728}$,
D. Ricci$^{720}$,
G. Scarpetta$^{702,722}$,
J. Skottfelt$^{710}$,
J. Southworth$^{725}$,
J. Surdej$^{720}$,
J. Tregloan-Reed$^{725}$,
J. Wambsganss$^{726}$,
O. Wertz$^{720}$\\
(The MiNDSTEp Consortium),\\
% uFUN -----------------------------
L. A. Almeida$^{u02}$,
V. Batista$^{u01}$,
G. Christie$^{u03}$,
D.L. DePoy$^{u04}$,
Subo Dong$^{u05}$,
B.S. Gaudi$^{u01}$,
C. Henderson$^{u01}$,
F. Jablonski$^{u02}$,
C.-U. Lee$^{713}$, 
J. McCormick$^{u07}$,
D. McGregor$^{u01}$,
D. Moorhouse$^{u08}$,
T. Natusch$^{u03,u09}$,
H. Ngan$^{u03}$,
S.-Y. Park$^{1}$,
R.W. Pogge$^{u01}$,
T.-G. Tan$^{u10}$,
G. Thornley$^{u08}$,
J.C. Yee$^{u01}$\\
(The $\mu$FUN Collaboration),\\
% PLANET ---------------------------
M.D. Albrow$^{P04}$,
E. Bachelet$^{P01}$,
J.-P. Beaulieu$^{P03}$,
S. Brillant$^{P08}$,
A. Cassan$^{P03}$,
A.A. Cole$^{P05}$,
E. Corrales$^{P03}$,
C. Coutures$^{P03}$,
S. Dieters$^{P05}$,
D. Dominis Prester$^{P09}$,
J. Donatowicz$^{P10}$,
J. Greenhill$^{P05}$,
D. Kubas$^{P08,P03}$,
J.-B. Marquette$^{P03}$,
J.W. Menzies$^{P02}$,
K.C. Sahu$^{P06}$,
M. Zub$^{726}$\\
(The PLANET Collaboration),\\
% ----------------------------------
}

\bigskip\bigskip
% --------------------------------------------------------------------------------------------------------------------------
\affil{$^{1}$Department of Physics, Institute for Astrophysics, Chungbuk National University, Cheongju 371-763, Korea}
% The OGLE Collaboration ---------------------------------------------------------------------------------------------------
\affil{$^{O01}$Warsaw University Observatory, Al. Ujazdowskie 4, 00-478 Warszawa, Poland}
\affil{$^{O02}$Universidad de Concepci\'{o}n, Departamento de Astronomia, Casilla 160--C, Concepci\'{o}n, Chile}
\affil{$^{O03}$Institute of Astronomy, University of Cambridge, Madingley Road, Cambridge CB3 0HA, United Kingdom}
% The MOA Collaboration ----------------------------------------------------------------------------------------------------
\affil{$^{M01}$Department of Earth and Space Science, Osaka University, Osaka 560-0043, Japan}
\affil{$^{M02}$Solar-Terrestrial Environment Laboratory, Nagoya University, Nagoya, 464-8601, Japan} 
\affil{$^{M03}$University of Notre Dame, Department of Physics, 225 Nieuwland Science Hall, Notre Dame, IN 46556-5670, USA} 
\affil{$^{M04}$Institute of Information and Mathematical Sciences, Massey University, Private Bag 102-904, North Shore Mail Centre, Auckland, New Zealand}
\affil{$^{M05}$Department of Physics, University of Auckland, Private Bag 92-019, Auckland 1001, New Zealand} 
\affil{$^{M06}$School of Chemical and Physical Sciences, Victoria University of Wellington, PO BOX 60, Wellington, New Zealand} 
\affil{$^{M07}$Okayama Astrophysical Observatory, NAOJ, Okayama 719-0232, Japan} 
\affil{$^{M08}$Nagano National College of Technology, Nagano 381-8550, Japan}
\affil{$^{M10}$Tokyo Metropolitan College of Aeronautics, Tokyo 116-8523, Japan} 
% The RoboNet Collaboration ------------------------------------------------------------------------------------------------ 
\affil{$^{R02}$Las Cumbres Observatory Global Telescope Network, 6740B Cortona Dr, Goleta, CA 93117, USA}
\affil{$^{R03}$School of Physics and Astronomy, Queen Mary University of London, Mile End Road, London, E1 4NS}
\affil{$^{R04}$European Southern Observatory, Karl-Schwarzschild-Str. 2, 85748 Garching bei M\"{u}nchen, Germany}
\affil{$^{R05}$Astrophysics Research Institute, Liverpool John Moores University, Liverpool CH41 1LD, UK}
\affil{$^{R06}$Max Planck Institute for Solar System Research, Max-Planck-Str. 2, 37191 Katlenburg-Lindau, Germany}
% The MiNDSTEp Consortium --------------------------------------------------------------------------------------------------
\affil{$^{701}$Qatar Foundation, P.O. Box 5825, Doha, Qatar}
\affil{$^{702}$Universit\`{a} degli Studi di Salerno, Dipartimento di Fisica "E.R. Caianiello", Via S. Allende, 84081 Baronissi (SA), Italy}
\affil{$^{703}$SUPA, University of St Andrews, School of Physics \& Astronomy, North Haugh, St Andrews, KY16 9SS, United Kingdom}
\affil{$^{704}$Deutsches SOFIA Institut, Universit\"{a}t Stuttgart, Pfaffenwaldring 31, 70569 Stuttgart, Germany}
\affil{$^{705}$SOFIA Science Center, NASA Ames Research Center, Mail Stop N211-3, Moffett Field CA 94035, United States of America}
\affil{$^{706}$Istituto Internazionale per gli Alti Studi Scientifici (IIASS), Vietri Sul Mare (SA), Italy}
\affil{$^{707}$Institut f\"{u}r Astrophysik, Georg-August-Universit\"{a}t, Friedrich-Hund-Platz 1, 37077 G\"{o}ttingen, Germany}
\affil{$^{708}$National Astronomical Observatories/Yunnan Observatory, Joint laboratory for Optical Astronomy, Chinese Academy of Sciences, Kunming 650011, People's Republic of China}
\affil{$^{709}$Department of Physics and Astronomy, Aarhus University, Ny Munkegade 120, 8000 {\AA}rhus C, Denmark}
\affil{$^{710}$Niels Bohr Institute, University of Copenhagen, Juliane Maries vej 30, 2100 Copenhagen, Denmark}
\affil{$^{711}$Centre for Star and Planet Formation, Geological Museum, {\O}ster Voldgade 5, 1350 Copenhagen, Denmark}
\affil{$^{712}$Armagh Observatory, College Hill, Armagh, BT61 9DG, Northern Ireland, United Kingdom}
\affil{$^{713}$Korea Astronomy and Space Science Institute, 776 Daedukdae-ro, Yuseong-gu, Daejeon 305-348, Republic of Korea}
\affil{$^{714}$Danmarks Tekniske Universitet, Institut for Rumforskning og -teknologi, Juliane Maries Vej 30, 2100 K{\o}benhavn, Denmark}
\affil{$^{715}$ESO Headquarters, Karl-Schwarzschild-Str. 2, 85748 Garching bei M\"{u}nchen, Germany}
\affil{$^{716}$Jodrell Bank Centre for Astrophysics, University of Manchester, Oxford Road,Manchester, M13 9PL, UK}
\affil{$^{717}$Max Planck Institute for Astronomy, K\"{o}nigstuhl 17, 69117 Heidelberg, Germany}
\affil{$^{719}$Department of Physics, Sharif University of Technology, P.~O.\ Box 11155--9161, Tehran, Iran}
\affil{$^{720}$Institut d'Astrophysique et de G\'{e}ophysique, All\'{e}e du 6 Ao\^{u}t 17, Sart Tilman, B\^{a}t.\ B5c, 4000 Li\`{e}ge, Belgium}
\affil{$^{722}$INFN, Gruppo Collegato di Salerno, Sezione di Napoli, Italy}
\affil{$^{725}$Astrophysics Group, Keele University, Staffordshire, ST5 5BG, United Kingdom}
\affil{$^{726}$Astronomisches Rechen-Institut, Zentrum f\"{u}r Astronomie der Universit\"{a}t Heidelberg (ZAH), M\"{o}nchhofstr.\ 12-14, 69120 Heidelberg, Germany}
\affil{$^{728}$Perimeter Institue for Theoretical Physics, 31 Caroline St. N., Waterloo, ON, N2L2Y5, Canada}
% The uFUN Collaboration ---------------------------------------------------------------------------------------------------
\affil{$^{u01}$Department of Astronomy, Ohio State University, 140 West 18th Avenue, Columbus, OH 43210, United States of America}
\affil{$^{u02}$Instituto Nacional de Pesquisas Espaciais, S\~{a}o Jos\'{e} dos Campos, SP, Brazil}
\affil{$^{u03}$Auckland Observatory, Auckland, New Zealand}
\affil{$^{u04}$Dept.\ of Physics, Texas A\&M University, College Station, TX, USA}
\affil{$^{u05}$Institute for Advanced Study, Einstein Drive, Princeton, NJ 08540, USA}
\affil{$^{u07}$Farm Cove Observatory, Centre for Backyard Astrophysics, Pakuranga, Auckland, New Zealand}
\affil{$^{u08}$Kumeu Observatory, Kumeu, New Zealand}
\affil{$^{u09}$AUT University, Auckland, New Zealand}
\affil{$^{u10}$Perth Exoplanet Survey Telescope, Perth, Australia}
% The PLANET Collaboration -------------------------------------------------------------------------------------------------
\affil{$^{P01}$IRAP, Universit\'e de Toulouse, CNRS, 14 Avenue Edouard Belin, 31400 Toulouse, France}
\affil{$^{P02}$South African Astronomical Observatory, P.O. Box 9 Observatory 7925, South Africa}
\affil{$^{P03}$UPMC-CNRS, UMR 7095, Institut d'Astrophysique de Paris, 98bis boulevard Arago, F-75014 Paris, France}
\affil{$^{P04}$University of Canterbury, Department of Physics and Astronomy, Private Bag 4800, Christchurch 8020, New Zealand}
\affil{$^{P05}$University of Tasmania, School of Mathematics and Physics, Private Bag 37, Hobart, TAS 7001, Australia}
\affil{$^{P06}$Space Telescope Science Institute, 3700 San Martin Drive, Baltimore, MD 21218, United States of America}
\affil{$^{P07}$Department of Physics, Massachussets Institute of Technology, 77 Mass. Ave., Cambridge, MA 02139, USA}
\affil{$^{P08}$European Southern Observatory, Casilla 19001, Vitacura 19, Santiago, Chile}
\affil{$^{P09}$Department of Physics, University of Rijeka, Omladinska 14, 51000 Rijeka, Croatia}
\affil{$^{P10}$Technische Universit\"{a}t Wien, Wieder Hauptst. 8-10, A-1040 Vienna, Austria}
% --------------------------------------------------------------------------------------------------------------------------
\affil{$^{\star}$Royal Society University Research Fellow}
\affil{$^{G1}$The OGLE Collaboration}
\affil{$^{G2}$The MOA Collaboration}
\affil{$^{G3}$The $\mu$FUN Collaboration}
\affil{$^{G4}$The RoboNet Collaboration}
\affil{$^{G5}$The MiNDSTEp Collaboration}
\affil{$^{G6}$The PLANET Collaboration}
% --------------------------------------------------------------------------------------------------------------------------
\affil{$^{S}$Corresponding author}
% --------------------------------------------------------------------------------------------------------------------------

\begin{abstract}
Despite astrophysical importance of binary star systems, detections are limited to those located in small ranges of separations, 
distances, and masses and thus it is necessary to use a variety of observational techniques for a complete view of stellar multiplicity 
across a broad range of physical parameters. In this paper, we report the detections and measurements of 2 binaries discovered from 
observations of microlensing events MOA-2011-BLG-090 and OGLE-2011-BLG-0417. Determinations of the binary masses are possible by 
simultaneously measuring the Einstein radius and the lens parallax. The measured masses of the binary components are 0.43 $M_{\odot}$ 
and 0.39 $M_{\odot}$ for MOA-2011-BLG-090 and 0.57 $M_{\odot}$ and 0.17 $M_{\odot}$ for OGLE-2011-BLG-0417 and thus both lens components 
of MOA-2011-BLG-090 and one component of OGLE-2011-BLG-0417 are M dwarfs, demonstrating the usefulness of microlensing in detecting 
binaries composed of low-mass components. From modeling of the light curves considering full Keplerian motion of the lens, we also 
measure the orbital parameters of the binaries. The blended light of OGLE-2011-BLG-0417 comes very likely from the lens itself, making 
it possible to check the microlensing orbital solution by follow-up radial-velocity observation. For both events, the caustic-crossing 
parts of the light curves, which are critical for determining the physical lens parameters, were resolved by high-cadence survey 
observations and thus it is expected that the number of microlensing binaries with measured physical parameters will increase in the future.
\end{abstract}


\section{Introduction}

Binary star systems are of astrophysical importance for various reasons. First, they compose an important portion of stars in the 
Galaxy \citep{abt76,abt83,duquennoy91} and thus theories about stellar formation and evolution should account for the binary nature 
of stars. Second, binary stars allow us to directly measure the masses of their component stars. The determined masses in turn allow 
other stellar parameters, such as radius and density, to be indirectly estimated. These physical parameters help us to understand the 
processes by which binary stars form \citep{goodwin07,burgasser07}. In particular, the separation and mass of a binary system tell us 
about the amount of angular momentum in the system. Because it is a conserved quantity, binaries with measured angular momentum give 
us important clues about the conditions under which the stars were formed.

Despite the importance, broad ranges of separations, distances, and component masses make it hard to detect and measure all binaries. 
Nearby systems with wide separations may be directly resolved using high-resolution imaging, while systems with small separations can 
be detected as eclipsing or spectroscopic binaries. However, binaries with intermediate separations are difficult to be detected by the 
conventional methods. In addition, it is difficult to detect binaries if they are located at large distances or either of the binary 
components is faint. As a result, samples are restricted to binaries in the solar neighborhood and are not complete down to low-mass 
stars. For a complete view of stellar multiplicity across a broad range of physical parameters, therefore, it is necessary to use a 
variety of observational techniques.

Gravitational microlensing can provide a complementary method that can detect and measure binaries that are difficult to be detected by 
other methods. Microlensing occurs when an astronomical object is closely aligned with a background star. The gravity of the intervening 
object (lens) causes deflection of the light from the background star (source), resulting in the brightening of the source star. If the lens 
is a single star, the light curve of the source star brightness is characterized by smooth rising and fall. However, if the lens is a 
binary, the light curve can be dramatically different, particularly for caustic-crossing events, which exhibit strong spikes in the light 
curve. Among caustic-crossing binary-lens events, those with long time scales are of special importance because it is often possible to 
determine the physical parameters of lenses (see more details in section 2). The binary separations for which caustic crossings are likely 
to occur are in the range of order AU, for which binaries are difficult to be detected by other methods. In addition, due to the nature of 
the lensing phenomenon that occurs regardless of the lens brightness, microlensing can provide an important channel to study binaries 
composed of low-mass stars.  Furthermore, most microlensing binaries are located at distances of order kpc and thus microlensing can 
expand the current binary sample throughout the Galaxy.

In this paper, we report the detections and measurements of 2 binaries discovered from observations of long time-scale caustic-crossing 
binary microlensing events MOA-2011-BLG-090 and OGLE-2011-BLG-0417. In \S 2, we describe the basic physics of binary lensing and the method to 
determine the physical parameters of binary lenses. In \S 3, we describe the choice of sample, observations of the events, and data reduction. 
In \S 4, we describe the procedure of modeling the observed light curves. In \S 5, we present the results from the analysis. We discuss about 
the findings and conclude in \S 6.

\section{LONG TIME-SCALE CAUSTIC-CROSSING EVENTS}

For a general lensing event, where a single star causes the brightening of a background source star, the magnification of the source star flux 
depends only on the projected separation between the source and the lens as
\begin{equation}
A={u^2+2\over u \sqrt{u^2+4}},
\label{eq1}
\end{equation}
where the separation $u$ is normalized in units of the angular Einstein radius of the lens, $\theta_{\rm E}$. For a uniform change of the lens-source 
separation, the light curve of a single-lens event is characterized by a smooth and symmetric shape. The normalized lens-source separation is related 
to the lensing parameters by
\begin{equation}
u=\left[ \left({t-t_0}\over{t_{\rm E}} \right)^2+u_0^2\right]^{1/2},
\label{eq2}
\end{equation}
where $t_{\rm E}$ represents the time scale for the lens to cross the Einstein radius (Einstein time scale), $t_0$ is the time of 
the closest lens-source approach, and $u_0$ is the lens-source separation at that moment. Among these lensing parameters $t_0$, $t_{\rm E}$, and $u_0$, 
the only quantity related to the physical parameters of the lens is the Einstein time scale. However, it results from the combination of the lens mass, 
distance, and transverse speed of the relative lens-source motion and thus the information about the lens from the time scale is highly degenerate.

When gravitational lensing is caused by a binary, the gravitational field is asymmetric and the resulting light curves can be dramatically different from that 
of a single lensing event \citep{mao91}. The most prominent feature of binary lensing that differentiates it from single lensing is a caustic. A set of caustics 
form a boundary of an envelope of rays as a curve of concentrated light. The gradient of magnification around the caustic is very large. As a result, 
the light curve of an event produced by the crossing of a source star over the caustic formed by a binary lens is characterized by sharp spikes occurring 
at the time of caustic crossings.

Caustic-crossing binary-lens events are useful because it is often possible to measure an additional lensing parameter appearing in the expression of the Einstein radius. 
This is possible because the caustic-crossing part of the light curve appears to be different for events associated with source stars of different sizes 
\citep{dominik95, gaudi99, gaudi02, pejcha09}. By measuring the deviation caused by this finite-source effect, it is possible to measure the source 
radius in units of the Einstein radius, $\rho_{\star}$ (normalized source radius). Then, combined with the information about the source angular size, 
$\theta_{\star}$, the Einstein radius is determined as $\theta_{\rm E}=\theta_{\star}/\rho_{\star}$. The Einstein radius is related to the mass, $M$, 
and distance to the lens, $D_{\rm L}$, by 
\begin{equation}
\theta_{\rm E}=(\kappa M \pi_{\rm rel})^{1/2};\qquad
\pi_{\rm rel}={\rm AU}\left( {1\over D_{\rm L}}-{1 \over D_{\rm S}}\right),
\label{eq3}
\end{equation}
where $\kappa=4G/({c^2}{\rm AU})$, $D_{\rm S}$ is the distance to the source, and $\pi_{\rm rel}$ represents the relative lens-source proper motion.
Unlike the Einstein time scale, the Einstein radius does not depend on the transverse speed of the lens-source motion and thus the physical parameters 
are less degenerate compared to the Einstein time scale.

Among caustic-crossing events, those with long time scales are of special interest because it is possible to completely resolve the degeneracy of the 
lens parameters and thus uniquely determine the mass and distance to the lens. This is possible because an additional lensing parameter of the lens 
parallax can be measured for these events. The lens parallax is defined as the ratio of Earth's orbit, i.e. 1 AU, to the physical Einstein radius, 
$r_{\rm E}=D_{\rm L}\theta_{\rm E}$, projected on the observer plane, i.e.
\begin{equation}
\pi_{\rm E}={\pi_{\rm rel}\over\theta_{\rm E}}.
\label{eq4}
\end{equation}
With simultaneous measurements of the Einstein radius and the lens parallax, the mass and distance to the lens are uniquely determined as 
\begin{equation}
M={\theta_{\rm E}\over \kappa \pi_{\rm E}}
\label{eq5}
\end{equation}
and
\begin{equation}
D_{\rm L}={{\rm AU}\over \pi_{\rm E}\theta_{\rm E}+\pi_{\rm S}};\qquad \pi_{\rm S}={{\rm AU}\over D_{\rm S}},
\label{eq6}
\end{equation}
respectively \citep{gould00}. The lens parallax is measured by analyzing deviations in lensing light curves caused by the deviation of 
the relative lens-source motion from a rectilinear one due to the change of the observer's position induced by the orbital motion of the Earth 
around the Sun \citep{gould92, refsdal66}. This deviation becomes important for long time-scale events, which endure for a significant fraction of 
the orbital motion of the Earth. Therefore, the probability of measuring the lens parallax is high for long time-scale events.

\section{SAMPLE AND OBSERVATIONS}

We searched for long time-scale caustic-crossing binary events among lensing events discovered in the 2011 microlensing observation season. 
We selected events to be analyzed based on the following criteria.
\begin{enumerate}
\item
The overall light curve was well covered with good photometry.
\item
At least one of caustic crossings was well resolved for the Einstein radius measurement.
\item
The time scale of an event should be long enough for the lens parallax measurement.
\end{enumerate}
From this search, we found 2 events of MOA-2011-BLG-090 and OGLE-2011-BLG-0417. Besides these events, there exist several other long 
time-scale caustic-crossing events, including MOA-2011-BLG-034, OGLE-2011-BLG-0307/MOA-2011-BLG-241, and MOA-2011-BLG-358/OGLE-2011-BLG-1132. 
We did not include MOA-2011-BLG-034 and MOA-2011-BLG-358/OGLE-2011-BLG-1132 in our analysis list because the coverage and photometry of the 
events are not good enough to determine the physical lens parameters by measuring subtle second-order effects in the lensing light curve. 
For OGLE-2011-BLG-0307/MOA-2011-BLG-241, the signal of the parallax effect was not strong enough to securely measure the physical parameters 
of the lens.

% Table 1 ----------------------------------------------------
\begin{deluxetable}{ll}
\tablecaption{Telescopes\label{table:one}}
\tablewidth{0pt}
\tablehead{
\multicolumn{1}{c}{event} &
\multicolumn{1}{c}{telescope}
}
\startdata
MOA-2011-BLG-090    & MOA MOA-II 1.8 m, New Zealand        \\
                    & OGLE Warsaw 1.3 m, Chile             \\
                    & $\mu$FUN CTIO/SMARTS2 1.3 m, Chile   \\
                    & $\mu$FUN PEST 0.3 m, Australia       \\
                    & MiNDSTEp Danish 1.54 m, Chile        \\
                    & RoboNet FTS 2.0 m, Australia         \\
OGLE-2011-BLG-0417  & OGLE Warsaw 1.3 m, Chile             \\
                    & $\mu$FUN CTIO/SMARTS2 1.3 m, Chile   \\ 
                    & $\mu$FUN Auckland 0.4 m, New Zealand \\
                    & $\mu$FUN FCO 0.36 m, New Zealand     \\ 
                    & $\mu$FUN Kumeu 0.36 m, New Zealand   \\
                    & $\mu$FUN OPD 0.6 m, Brazil           \\
                    & PLANET Canopus 1.0 m, Australia      \\
                    & PLANET SAAO 1.0 m, South Africa      \\
                    & MiNDSTEp Danish 1.54 m, Chile        \\
                    & RoboNet FTN 2.0 m, Hawaii            \\
                    & RoboNet LT 2.0 m, Spain         
\enddata  
%\tablecomments{ 
%}
\end{deluxetable}
% ------------------------------------------------------------

The events MOA-2011-BLG-090 and OGLE-2011-BLG-0417 were observed by the microlensing experiments that are being conducted toward Galactic bulge 
fields by 6 different groups including MOA, OGLE, $\mu$FUN, PLANET, RoboNet, and MiNDSTEp. Among them, the MOA and OGLE collaborations 
are conducting survey observations for which the primary goal is to detect a maximum number of lensing events by monitoring a large area of sky. 
The $\mu$FUN, PLANET, RoboNet, and MiNDSTEp groups are conducting follow-up observations of events detected from survey observations. The events 
were observed by using 12 telescopes located in 3 different continents in the Southern Hemisphere. In Table \ref{table:one}, we list the telescopes 
used for the observation.

Reduction of the data was conducted by using photometry codes developed by the individual groups. The OGLE and MOA data were reduced by photometry 
codes developed by \citet{udalski03} and \citet{bond01}, respectively, which are based on the Difference Image Analysis method \citep{alard98}. The 
$\mu$FUN data were processed using a pipeline based on the DoPHOT software \citep{schechter93}. For PLANET and MiNDSTEp data, a pipeline based 
on the pySIS software \citep{albrow09} is used. For RoboNet data, the DanDIA pipeline \citep{bramich08} is used.

To standardize error bars of data estimated from different observatories, we re-scaled them so that $\chi^2$ per degree of freedom becomes unity 
for the data set of each observatory, where $\chi^2$ is computed based on the best-fit model. For the data sets used for modeling, we eliminate 
data points with very large photometric uncertainties and those lying beyond $3\sigma$ from the best-fit model.

\begin{figure}[ht]
\epsscale{1.1}
\plotone{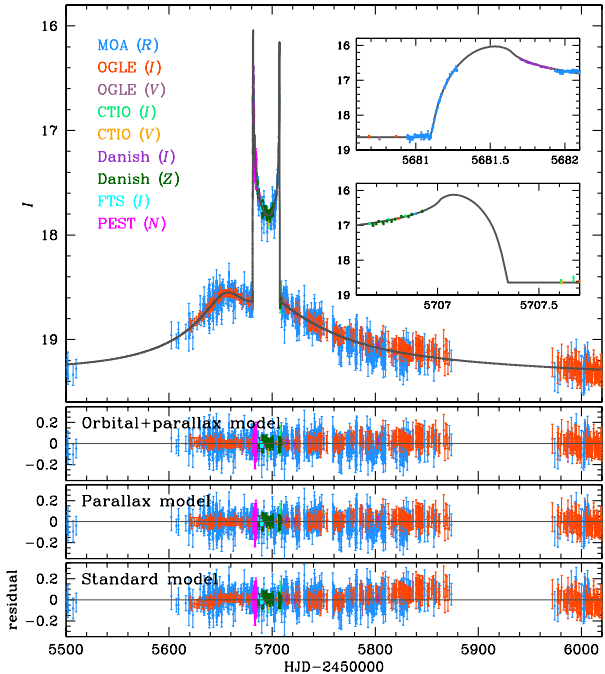}
\caption{\label{fig:one}
Light curve of MOA-2011-BLG-090. The notation in the parentheses after each observatory denotes the 
passband of observation, where $N$ denotes that no filter was used. The insets in the upper panel show 
the enlargement of the caustic crossing parts of the light curve. The lower 3 panels show the residuals 
from the 3 tested models. See section 4 for more details about the individual models. 
}\end{figure}

\begin{figure}[ht]
\epsscale{1.1}
\plotone{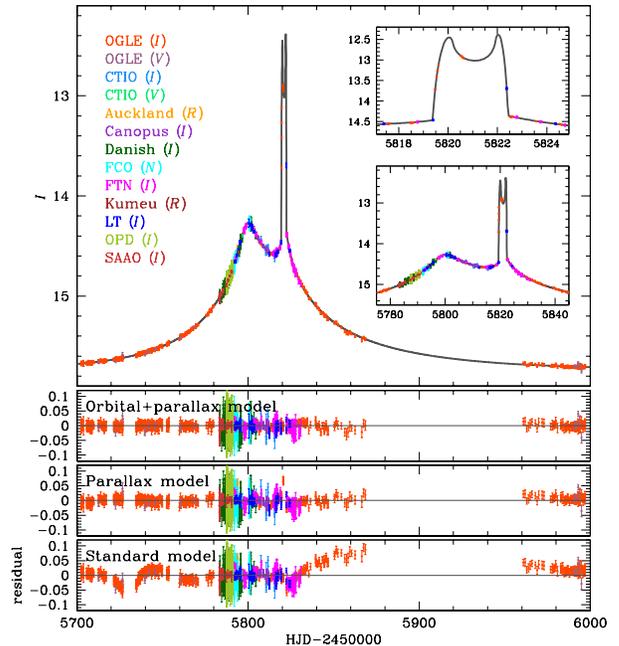}
\caption{\label{fig:two}
Light curve of OGLE-2011-BLG-0417. Notations are same as in Fig 1.
}\end{figure}

We present the light curves of events in Figure \ref{fig:one} and \ref{fig:two}. To be noted is that the overall light curves including caustic 
crossings of both events are well covered by survey observations. This demonstrates that the observational cadence of survey experiments is now 
high enough to characterize lensing events based on their own data. The parts of light curves with 2455880 < HJD < 2455960 were not covered because the 
Galactic bulge field could not be observed. Although not included in the selection criteria, both events showed a common bump to those involved 
with caustic crossings: at HJD$\sim$2455655 for MOA-2011-BLG-090 and at HJD$\sim$2455800 for OGLE-2011-BLG-0417. These bumps were produced during 
the approach of the source trajectory close to a cusp of a caustic. \citet{an01} pointed out that such triple-peak features help to better measure 
the lens parallax.

\section{MODELING LIGHT CURVES}

In modeling the light curve of each event, we search for a solution of lensing parameters that best characterizes the observed light curve. 
Describing the basic feature of a binary-lens light curve requires 6 parameters including the 3 single-lensing parameters $t_0$, $u_0$, and 
$t_{\rm E}$. The 3 additional binary-lensing parameters include the mass ratio between the lens components, $q$, the projected separation in 
units of the Einstein radius, $s_{\perp}$, and the angle between the source trajectory and the binary axis, $\alpha$ (trajectory angle).

In addition to the basic binary lensing parameters, it is required to include additional parameters to precisely describe detailed features caused 
by various second-order effects. The first such effect is related to the finite size of the source star. This finite-source effect becomes important 
when the source is located at a position where the gradient of magnification is very high and thus different parts of the source surface experience 
different amounts of magnification. For binary-lens events, this happens when the source approaches or crosses the caustic around which the gradient 
of magnification is very high. To describe the light curve variation caused by the finite-source effect, it is necessary to include an additional 
parameter of the normalized source radius, $\rho_{\star}$.

% Figure 1 ----------------------------------------------------------------------------------------
% --------------------------------------------------------------------------------------------------

For long time-scale events, such as those analyzed in this work, it is required to additionally consider the parallax effect. Consideration of 
the parallax effect in modeling requires to include 2 additional parameters $\pi_{{\rm E},N}$, and $\pi_{{\rm E},E}$, which represent the two 
components of the lens parallax vector $\pivec_{\rm E}$ projected on the sky along the north and east equatorial coordinates, respectively. 
The direction of the lens parallax vector corresponds to the relative lens-source relative motion in the frame of the Earth at a specific time 
of the event. The size of the vector corresponds to the ratio of the Earth's orbit to the Einstein radius projected on the observer's plane, 
i.e.\ $\pi_{\rm E}={\rm AU}/[r_{\rm E}D_{\rm S}/(D_{\rm S}-D_{\rm L})]$, where $r_{\rm E}=D_{\rm L}\theta_{\rm E}$ is the physical size of the 
Einstein radius.

Another effect to be considered for long time-scale events is the orbital motion of the lens \citep{dominik98, ioka99, albrow00, penny11, 
shin11, skowron11}.  The lens orbital motion affects lensing light curves in two different ways. First, it causes to change the binary 
separation and thus the magnification pattern.  Second, it also causes the binary axis to rotate with respect to the source trajectory. 
In order to fully account for the lens orbital motion, 4 additional parameters are needed.  The first two of these parameters are 
$ds_{\perp}/dt$ and $d\alpha/dt$, which represent the change rates of the projected binary separation and the trajectory angle, respectively. 
The other two orbital parameters are $s_{\parallel}$ and $ds_{\parallel}/dt$, where $s_{\parallel}$ represents the line-of-sight separation 
between the binary components in units of $\theta_{\rm E}$ and $ds_{\parallel}/dt$ represents its rate of change. For a full description of 
the orbital lensing parameters, see the summary in the Appendix of \citet{skowron11}. The deviation in a lensing light curve affected by the 
orbital effect is smooth and long lasting and thus is similar to the deviation induced by the parallax effect. This implies that if the orbital 
effect is not considered, the measured lens parallax and the resulting lens parameters might be erroneous. Therefore, considering the orbital 
effect is important not only for constraining the orbital motion of the lens but also for precisely determining the physical parameters of the lens.

% Figure 2 -----------------------------------------------------------------------------------------
% --------------------------------------------------------------------------------------------------

With all these parameters, we test three different models.
In the first model, we fit the light curve with standard binary lensing parameters considering the finite-source effect (standard model).
In the second model, we additionally consider the parallax effect (parallax model).  Finally, we take the orbital effect into consideration 
as well (orbital model).  When the source trajectory is a straight line, the two light curves resulting from source trajectories with 
positive ($+u_0$) and negative ($-u_0$) impact parameters are identical due to the symmetry of the magnification pattern with respect to 
the binary axis. When either the parallax or the orbital effect is considered, on the other hand, the source trajectory deviates from a 
straight line and thus the light curves with $+u_0$ and $-u_0$ are different from each other. We, therefore, consider both the $+u_0$ and 
$-u_0$ cases for each of the models considering the parallax and orbital effects.

In modeling, the best-fit solution is obtained by minimizing $\chi^2$ in the parameter space. We conduct this in 3 stages. In the first stage, grid 
searches are conducted over the space of a subset of parameters and the remaining parameters are searched by using a downhill approach \citep{dong06}. We then identify 
local minima in the grid-parameter space by inspecting the $\chi^2$ distribution. In the second stage, we investigate the individual local minima by allowing 
the grid parameters to vary and find the exact location of each local minimum. In the final stage, we choose the best-fit solution by comparing $\chi^2$ 
values of the individual local minima. This multiple stage procedure is needed for thorough investigation of possible degeneracy of solutions. We choose of 
$s_{\perp}$, $q$, and $\alpha$ as the grid parameters because they are related to the light curve features in a complex way such that a small change in the 
values of the parameters can lead to dramatic changes in the resulting light curve. On the other hand, the other parameters are more directly related to the 
light curve features and thus they are searched for by using a downhill approach. For the $\chi^2$ minimization in the downhill approach, we use 
the Markov Chain Monte Carlo (MCMC) method. Once a solution of the parameters is found, we estimate the uncertainties of the individual parameters based on 
the chain of solutions obtained from MCMC runs.

% Figure 3 -----------------------------------------------------------------------------------------
\begin{figure}[ht]
\epsscale{1.1}
\plotone{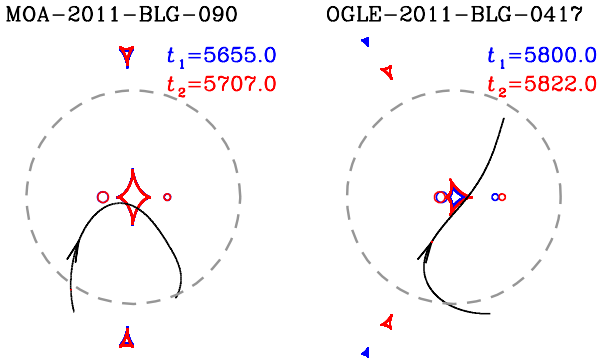}
\caption{\label{fig:three}
The caustic geometries and source trajectories of the best-fit models for MOA-2011-BLG-090 (left panel) 
and OGLE-2011-BLG-0417 (right panel). The small closed figures composed of concave curves represent the 
caustics. The small open circles represent the lens positions. We note that the lens positions and the 
resulting caustic vary in time due to the orbital motion of the lens. We mark 2 sets of lens positions 
and caustics at 2 different times of $t_1$  and $t_2$. We also note that the source trajectory, the curve 
with an arrow, is curved due to the combination of the parallax and orbital effects. The source trajectory 
is presented so that the binary axis is set along the horizontal axis. The dashed circles represent the 
Einstein rings corresponding to the total mass of the binary lenses.
}\end{figure}
% --------------------------------------------------------------------------------------------------

% Table 2 ---------------------------------------------------------------------------------------------
\begin{deluxetable}{ccc}
\tablecaption{Limb-darkening coefficients\label{table:two}}
\tablewidth{0pt}
\tablehead{
\multicolumn{1}{c}{quantity} &
\multicolumn{1}{c}{MOA-2011-BLG-090} &
\multicolumn{1}{c}{OGLE-2011-BLG-0417}
}
\startdata
$\Gamma_{V}$                      & 0.52  & 0.71  \\
$\Gamma_{R}$                      & 0.45  & 0.61  \\
$\Gamma_{I}$                      & 0.37  & 0.51  \\
source type                       & FV    & KIII  \\
$T_{\rm eff}$ (K)                 & 6650  & 4660  \\
$v_{\rm turb}$ ($\rm km\ s^{-1}$) & 2     & 2     \\
$\log{g}$ ($\rm cm\ s^{-2}$)      & 4.5   & 2.5  
\enddata  
%\tablecomments{ 
%}
\end{deluxetable}
% -----------------------------------------------------------------------------------------------------

To compute lensing magnifications affected by the finite-source effect, we use the ray-shooting method. \citep{schneider86, kayser86, wambsganss97}. 
In this method, rays are uniformly shot from the image plane, bent according to the lens equation, and land on the source plane. Then, a finite magnification 
is computed by comparing the number densities of rays on the image and source planes. Precise computation of finite magnifications by using this numerical 
technique requires a large number of rays and thus demands heavy computation. To minimize computation, we limit finite-magnification computation by using 
the ray-shooting method only when the lens is very close to caustics. In the adjacent region, we use an analytic hexadecapole approximation 
\citep{pejcha09, gould08}. In the region with large enough distances from caustics, we use a point-source magnifications.

% Figure 4 -----------------------------------------------------------------------------------------
\begin{figure}[ht]
\epsscale{1.1}
\plotone{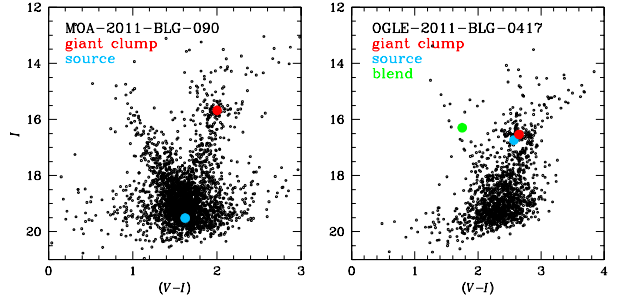}
\caption{\label{fig:four}
OGLE color-magnitude diagrams of stars in the fields where the lensing events MOA-2011-BLG-090 (left panel) 
and OGLE-2011-BLG-0417 (right panel) occurred. The red and blue dots represent the centroid of the giant 
clump and the location of the source star for each event, respectively. For OGLE-2011-BLG-0417, we also 
mark the position of the blended star.
}\end{figure}
% ---------------------------------------------------------------------------------------------------

In the finite magnification computation, we consider the variation of the magnification caused by the limb-darkening of the source star's surface. We model 
the surface brightness profile of a source star as
\begin{equation}
S_{\lambda}={F_{\lambda} \over{\pi{\theta_{\star}}^2}} \left[ 1-\Gamma_{\lambda} \left( 1- {3\over{2}} \cos\psi  \right) \right],
\end{equation}
where $\Gamma_{\lambda}$ is the linear limb-darkening coefficients, $F_{\lambda}$ is the source star flux, and $\psi$ is the angle between the normal to the source 
star's surface and the line of sight toward the star. The limb-darkening coefficients are set based the source type that is determined on the basis of the color 
and magnitude of the source. In Table \ref{table:two}, we present the used limb-darkening coefficients, the corresponding source types, and the measured de-reddened 
color along with the assumed values of the effective temperature, $T_{\rm eff}$, the surface turbulence velocity, $v_{\rm turb}$, and the surface gravity, 
$\log{g}$. For both events, we assume a solar metallicity.

\section{RESULTS}

In Table \ref{table:three}, we present the solutions of parameters for the tested models. The best-fit model light curves are drawn on the top of the observed 
light curves in Figures \ref{fig:one} and \ref{fig:two}. In Figure \ref{fig:three}, we present the geometry of the lens systems where the source 
trajectory with respect to the caustic and the locations of the lens components are marked. We note that the source trajectories are curved due to the 
combination of the parallax and orbital effects. We also note that the positions of the lens components and the corresponding caustics change in time due 
to the orbital motion and thus we present caustics at 2 different moments that are marked in Figure \ref{fig:three}. These moments correspond to those of 
characteristic features on the light curve such as the peak involved during a cusp approach or a caustic crossing. To better show the differences in the fit 
between different models, we present the residuals of data from the best-fits of the individual models. For a close-up view of the caustic-crossing 
parts of the light curves, we also present enlargement of the light curve.

For both events, the parallax and orbital effects are detected with significant statistical confidence levels. It is found that inclusion of the second-order 
effects of the parallax and orbital motions improves the fits with $\Delta\chi^2=571$ and $2680$ for MOA-2011-BLG-090 and OGLE-2011-BLG-0417, respectively. 
To be noted is that the orbital effect is considerable for OGLE-2011-BLG-0417 and thus the difference between the values of the lens parallax measured with 
($\pi_{\rm E}=0.40$) and without ($\pi_{\rm E}=0.17$) considering the orbital effect is substantial. Since the lens parallax is directly related to the 
physical parameters of the lens, this implies that considering the orbital motion of the binary lens is important for the accurate measurement of the lens 
parallax and thus the physical parameters.

% Table 3 ---------------------------------------------------------------------------------------------------------------------------------------------------------------
\begin{deluxetable*}{lrrrrrr}
\tablecaption{Lensing Parameters\label{table:three}}
\tablewidth{0pt}
\tablehead{
\multicolumn{1}{l}{parameters} &
\multicolumn{3}{c}{MOA-2011-BLG-090} &
\multicolumn{3}{c}{OGLE-2011-BLG-0417} \\
\multicolumn{1}{c}{} &
\multicolumn{3}{c}{model} &
\multicolumn{3}{c}{model} \\
\multicolumn{1}{c}{} &
\multicolumn{1}{c}{standard} &
\multicolumn{1}{c}{parallax} &
\multicolumn{1}{c}{orbital+parallax} &
\multicolumn{1}{c}{standard} &
\multicolumn{1}{c}{parallax} &
\multicolumn{1}{c}{orbital+parallax}
}
\startdata
$\chi^2/{\rm dof}$                  & 5207/5164          &  4718/5162          &  4636/5158          &  4415/2627          &  2391/2625          &  1735/2621          \\
$t_0$ (HJD')                        & 5688.331$\pm$0.121 &  5691.563$\pm$0.187 &  5690.409$\pm$0.110 &  5817.302$\pm$0.018 &  5815.867$\pm$0.030 &  5813.306$\pm$0.059 \\
$u_0$                               & 0.3307$\pm$0.0038  & -0.0613$\pm$0.0008  & -0.0785$\pm$0.0005  &  0.1125$\pm$0.0001  & -0.0971$\pm$0.0003  & -0.0992$\pm$0.0005  \\
$t_{\rm E}$ (days)                  & 94.10$\pm$0.71     &  279.88$\pm$0.27    &  220.40$\pm$0.21    &  60.74$\pm$0.08     &  79.59$\pm$0.36     &  92.26$\pm$0.37     \\
$s_\perp$                           & 0.981$\pm$0.002    &  0.536$\pm$0.002    &  0.606$\pm$0.001    &  0.601$\pm$0.001    &  0.574$\pm$0.001    &  0.577$\pm$0.001    \\
$q$                                 & 0.611$\pm$0.005    &  1.108$\pm$0.026    &  0.892$\pm$0.014    &  0.402$\pm$0.002    &  0.287$\pm$0.002    &  0.292$\pm$0.002    \\
$\alpha$ (rad)                      & -0.181$\pm$0.004   &  0.373$\pm$0.005    &  0.317$\pm$0.006    &  1.030$\pm$0.002    & -0.951$\pm$0.002    & -0.850$\pm$0.004    \\
$\rho_\star$ ($10^{-3}$)            & 2.89$\pm$0.03      &  0.54$\pm$0.01      &  0.78$\pm$0.01      &  3.17$\pm$0.01      &  2.38$\pm$0.02      &  2.29$\pm$0.02      \\
$\pi_{{\rm E},N}$                   & --                 &  0.205$\pm$0.003    &  0.159$\pm$0.003    & --                  &  0.125$\pm$0.004    &  0.375$\pm$0.015    \\
$\pi_{{\rm E},E}$                   & --                 & -0.071$\pm$0.005    & -0.022$\pm$0.004    & --                  & -0.111$\pm$0.005    & -0.133$\pm$0.003    \\
$ds_\perp/dt$ (${\rm yr}^{-1}$)     & --                 & --                  & -0.031$\pm$0.007    & --                  & --                  &  1.314$\pm$0.023    \\
$d\alpha/dt$  (${\rm yr}^{-1}$)     & --                 & --                  &  1.066$\pm$0.005    & --                  & --                  &  1.168$\pm$0.076    \\
$s_\parallel$                       & --                 & --                  &  0.137$\pm$0.008    & --                  & --                  &  0.467$\pm$0.020    \\
$ds_\parallel/dt$ (${\rm yr}^{-1}$) & --                 & --                  & -0.784$\pm$0.008    & --                  & --                  & -0.192$\pm$0.036     
\enddata                             
\tablecomments{ 
HJD'=HJD-2450000.
}
\end{deluxetable*}
% -----------------------------------------------------------------------------------------------------------------------------------------------------------------------

% Table 4 ---------------------------------------------------------------------------------------------
\begin{deluxetable}{lrr}
\tablecaption{Physical Lens Parameters\label{table:four}}
\tablewidth{0pt}
\tablehead{
\multicolumn{1}{l}{parameter} &
\multicolumn{1}{c}{MOA-2011-BLG-090} &
\multicolumn{1}{c}{OGLE-2011-BLG-0417}
}
\startdata
$M_{\rm tot}$ ($M_\odot$) &   0.82$\pm$0.02  &   0.74$\pm$0.03  \\
$M_{\rm 1}$ ($M_\odot$)   &   0.43$\pm$0.01  &   0.57$\pm$0.02  \\
$M_{\rm 2}$ ($M_\odot$)   &   0.39$\pm$0.01  &   0.17$\pm$0.01  \\
$\theta_{\rm E}$ (mas)    &   1.06$\pm$0.01  &   2.44$\pm$0.02  \\
$\mu$ (mas yr$^{-1}$)     &   1.76$\pm$0.02  &   9.66$\pm$0.07  \\
$D_{\rm L}$ (kpc)         &   3.26$\pm$0.05  &   0.89$\pm$0.03  \\
$a$ (AU)                  &   1.79$\pm$0.02  &   1.15$\pm$0.04  \\
$P$ (yr)                  &   2.65$\pm$0.04  &   1.44$\pm$0.06  \\
$e$                       &   0.28$\pm$0.01  &   0.68$\pm$0.02  \\
$i$ ($\deg$)              & 129.43$\pm$0.33  & 116.95$\pm$1.04
\enddata  
\tablecomments{ 
$M_{\rm tot}$: total mass of the binary, $M_{\rm 1}$ and $M_{\rm 2}$: masses of the binary components, 
$\theta_{\rm E}$: angular Einstein radius, $\mu$: relative lens-source proper motion, 
$D_{\rm L}$: distance to the lens,  $a$: semi-major axis, $P$: orbital period, $e$: eccentricity, 
$i$: inclination of the orbit.
}
\end{deluxetable}
% -----------------------------------------------------------------------------------------------------

The finite-source effect is also clearly detected and the normalized source radii are precisely measured for both events. To obtain the Einstein 
radius from the measured normalized source radius, $\rho_{\star}$, additional information about the source star is needed. We obtain this 
information by first locating the source star on the color-magnitude diagram of stars in the field and then calibrating the source brightness 
and color by using the centroid of the giant clump as a reference under the assumption that the source and clump giants experience the same amount 
of extinction and reddening \citep{yoo04}. The measured $V/I$ colors are then translated into $V/K$ color by using the $V/I-V/K$ relations of 
\citet{bessell88} and the angular source radius is obtained by using the $V/K$ color and the angular radius given by \citet{kervella04}. In 
Figure \ref{fig:four}, we present the color-magnitudes of field stars constructed based on OGLE data and the locations of the source star. 
We find that the source star is a F-type main-sequence star with a de-reddened color of $(V-I)_0=0.68$ for MOA-2011-BLG-090 and a K-type giant 
with $(V-I)_0=0.98$ for OGLE-2011-BLG-0417. Here we assume that the de-reddened color and absolute magnitude of the giant clump centroid are 
$(V-I)_{0,c}=1.06$ and $M_{I,c}=-0.23$ \citep{stanek98}, respectively. The mean distances to clump stars of $\sim$7200 pc for MOA-2011-BLG-090 
and $\sim$7900 pc for OGLE-2011-BLG-0417 are estimated based on the Galactic model of \citet{han95}. The measured Einstein radii of the individual 
events are presented in Table \ref{table:four}. Also presented are the relative lens-source proper motions as determined by $\mu=\theta_{\rm E}/t_{\rm E}$.

With the measured lens parallax and the Einstein radius, the mass and distance to the lens of each event are determined by using the relations 
(\ref{eq5}) and (\ref{eq6}). The measured masses of the binary components are 0.43 $M_{\odot}$ and 0.39 $M_{\odot}$ for MOA-2011-BLG-090 and 
0.57 $M_{\odot}$ and 0.17 $M_{\odot}$ for OGLE-2011-BLG-0417. It is to be noted that both lens components of MOA-2011-BLG-090 and one component 
of OGLE-2011-BLG-0417 are M dwarfs which are difficult to be detected by other conventional methods due to their faintness. 
It is found that the lenses are located at distances $D_{\rm L}\sim3.3$ kpc and $0.9$ kpc for MOA-2011-BLG-090 and OGLE-2011-BLG-0417, respectively.

\begin{figure}[ht]
\epsscale{1.05}
\plotone{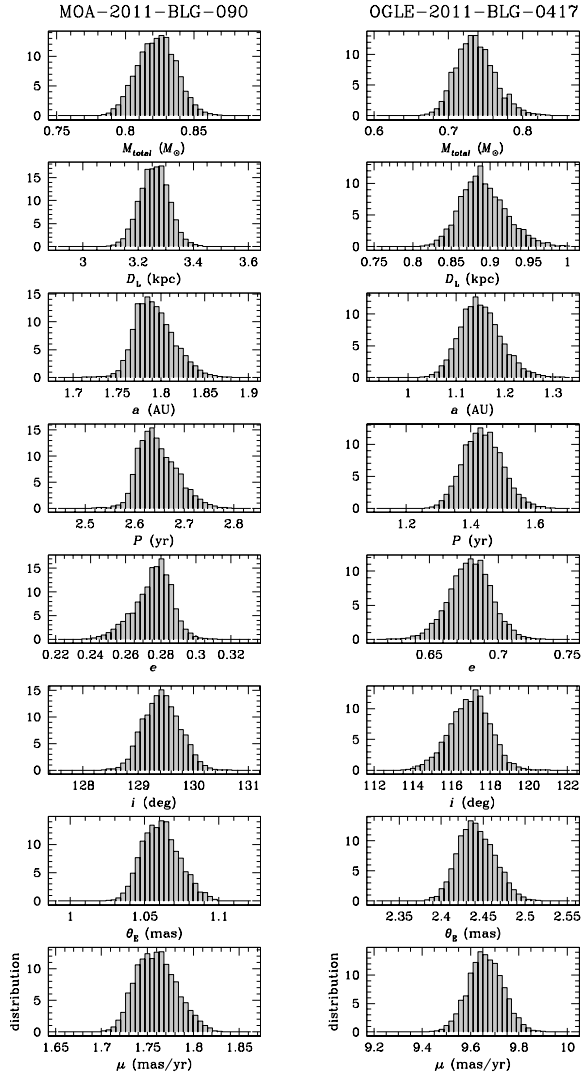}
\caption{\label{fig:five}
Distributions of the physical and orbital parameters of the binary lens systems for MOA-2011-BLG-090 
(left panels) and OGLE-2011-BLG-0417 (right panels).
}\end{figure}

Since full Keplerian motion of the binary lens is considered in our modeling, we also determine the orbital parameters of the semi-major axis 
$a$, period $P$, eccentricity $e$, and inclination $i$. We find that the binary lens components of MOA-2011-BLG-090 are orbiting each other 
with a semi-major axis of $\sim1.8$ AU and an orbital period of $\sim2.7$ yrs. For OGLE-2011-BLG-0417, the semi-major axis and the orbital period 
of the binary lens are $\sim1.2$ AU and $\sim1.4$ yrs, respectively. In Figure \ref{fig:five}, we present the distributions of the physical and 
orbital parameters constructed based on the MCMC chains. In Table \ref{table:four}, we summarize the measured parameters of the binary lenses 
for both events. We note that the uncertainties of the parameters are based on the standard deviations of the MCMC distributions.

We note that the blended light of OGLE-2011-BLG-0417 comes very likely from the lens itself, implying that the lens can be directly observed. 
Based on the measured mass of 0.57 $M_{\odot}$, the primary of the binary lens corresponds to a late K-type main sequence star with an absolute 
magnitude and a de-reddened color of $M_I\sim6.0$ and $(V-I)_0\sim1.5$, respectively. Considering the distance of 0.89 kpc and assuming an extinction 
of $A_I\sim0.5$ and the color index of $E(V-I)\sim0.3$, the apparent brightness and color of the lens correspond to $I\sim16.2$ and $(V-I)\sim1.8$, 
respectively. These values match very well with the location of the blend marked on the right panel of Figure \ref{fig:four}, implying that the blend 
is very likely the lens. The visibility of the lens signifies this event because it is possible to check the microlensing orbital solution by 
spectroscopic radial-velocity observation.

% Figure 5 -----------------------------------------------------------------------------------------
% ---------------------------------------------------------------------------------------------------

\section{DISCUSSION AND CONCLUSION}

We reported detections and measurements of 2 binaries discovered from observations of microlensing events MOA-2011-BLG-090 and OGLE-2011-BLG-0417. 
The light curves of the events have common characteristics of long durations with caustic-crossing features, which enabled to determine the physical 
parameters of the lenses. It was found that both lens components of MOA-2011-BLG-090 and one component of OGLE-2011-BLG-0417 are M dwarfs.  Therefore, 
the discovered microlensing binaries demonstrate the usefulness of gravitational lensing in detecting and characterizing binaries composed of low-mass 
stars. By considering full Keplerian binary motion, we also determined the orbital parameters of the binaries. For OGLE-2011-BLG-0417, the blended 
light comes very likely from the lens itself and thus it would be possible to check the orbital solution from follow-up radial-velocity observation.

Studies of M dwarfs are important not only because they are the most abundant stars in the Milky Way but also they form a link between solar-type stars 
and brown dwarfs; two mass regimes that might exhibit very different multiplicity characteristics. Precise knowledge of multiplicity characteristics and 
how they change in this transitional mass region provide constraints on low-mass star and brown dwarf formation \citep{goodwin07, burgasser07}. Despite 
the importance of M-dwarf binaries, only a few measurements of the binary fraction and distribution of low-mass stars have been made, e.g., \citet{delfosse04}, 
and the samples are restricted to only binaries in the solar neighborhood.  As a result, there are still large uncertainties about their basic physical properties 
as well as their formation environment. Considering the rapid improvement of lensing surveys both in equipment and strategy, it is expected that the number 
of microlensing binaries with measured physical parameters will increase rapidly. This will contribute to the complete view of stellar multiplicity across 
a wide range of binary parameters.

\acknowledgments 
Work by CH was supported by Creative Research Initiative Program 
(2009-0081561) of National Research Foundation of Korea.
The OGLE project has received funding from the European Research Council 
under the European Community's Seventh Framework Programme 
(FP7/2007-2013) / ERC grant agreement no. 246678. The MOA experiment was 
supported by grants JSPS22403003 and JSPS23340064. 
TS was supported by the grant JSPS23340044.
Y. Muraki acknowledges support from JSPS grants JSPS23540339 and JSPS19340058.
The MiNDSTEp monitoring campaign is powered by ARTEMiS
(Automated Terrestrial Exoplanet Microlensing Search; Dominik et al.
2008, AN 329, 248). MH acknowledges support by the German Research
Foundation (DFG). DR (boursier FRIA) and J. Surdej acknowledge support
from the Communaut\'{e} fran\c{c}aise de Belgique Actions de recherche
concert\'{e}es -- Acad\'{e}mie universitaire Wallonie-Europe.
The RoboNet team is supported by the Qatar Foundation through QNRF grant NPRP-09-476-1-78.
CS received funding from the European Union Seventh Framework Programme (FPT/2007-2013) under grant agreement 268421.
This work is based in part on data collected by MiNDSTEp with the Danish 1.54 m telescope at the ESO La Silla Observatory. 
The Danish 1.54 m telescope is operated based on a grant from the Danish Natural Science Foundation (FNU). 
A. Gould and B.S. Gaudi acknowledge support from NSF AST-1103471.
B.S. Gaudi, A. Gould, and R.W. Pogge acknowledge support from NASA grant NNG04GL51G.
Work by J.C. Yee is supported by a National Science Foundation Graduate Research Fellowship under Grant No. 2009068160.
S. Dong's research was performed under contract with the California Institute of Technology (Caltech) funded by NASA through the Sagan Fellowship Program.
Research by TCH was carried out under the KRCF Young Scientist Research Fellowship Program. 
TCH and CUL acknowledge the support of Korea Astronomy and Space Science Institute (KASI) grant 2012-1-410-02.

\end{document}